\documentclass[aps,prb,showpacs,preprintnumbers,twocolumn,superscriptaddress]{revtex4-2}
\usepackage{amsmath,amssymb}
\usepackage{bm}
\usepackage{tipa}
\usepackage{upgreek}
\usepackage{comment}
\usepackage{mathrsfs}
\usepackage{graphicx}
\usepackage{braket}
\usepackage{enumitem}
\usepackage{mathbbol}
\usepackage{booktabs}
\usepackage{gensymb}
\usepackage[normalem]{ulem}
\usepackage{color}
\usepackage[colorlinks,bookmarks=true,citecolor=blue,linkcolor=red,urlcolor=blue]{hyperref}
\usepackage{hyperref}
\renewcommand{\vec}[1]{\mathbf{#1}}
\usepackage{pifont}

\begin{document}

    \title{Sensing the binding and unbinding of anyons at impurities}
\author{Glenn Wagner}
\affiliation{Institute for Theoretical Physics, ETH Z{\"u}rich, 8093 Z{\"u}rich, Switzerland}
\author{Titus Neupert}
\affiliation{Department of Physics, University of Z{\"u}rich, Winterthurerstrasse 190, 8057 Z{\"u}rich, Switzerland}

	\begin{abstract}
    Anyons are quasiparticles with fractional charge and statistics that arise in strongly correlated two-dimensional systems such as the fractional quantum Hall (FQH) effect and fractional Chern insulators (FCI). Interactions between anyons can lead to emergent phenomena, such as anyon superconductivity as well as anyon condensation which allows for a hierarchical construction of quantum Hall states. In this work, we study how quasihole anyons in a $\nu=1/3$ Laughlin fractional quantum Hall state can be bound together by a sufficiently strong attractive impurity potential. The competition between the repulsive interaction between the quasiholes themselves and the attractive interaction between the quasiholes and the impurity leads to states with different numbers of quasiholes bound to the impurity. Tuning the chemical potential via gating while remaining within a quantum Hall plateau changes the number of quasiholes bound to the impurity.  We propose methods for studying these states experimentally, for example using scanning tunneling microscopy and exciton spectroscopy. While the impurities in traditional platforms such as GaAs heterostructures are typically too weak to observe the binding of anyons, the recently discovered zero-field fractional Chern insulators in twisted MoTe$_2$ offer a platform which may realize the strong-impurity regime. 
	\end{abstract}
	
 	\maketitle

\section{Introduction}

Impurities can be used to probe quantum systems. One of the most famous examples thereof are the nitrogen-vacancy centers in diamond \cite{Shirhagl2014}. Impurities and defects can also be used to probe topological states, such as in Shiba states to detect topological superconductivity \cite{Sau2013,Soldini2023} or step edges to probe topological crystalline insulators \cite{Sessi2016,Wagner2023,Wagner2025}. Magnetic impurities have also been proposed as a tool to detect the notoriously hard-to-diagnose quantum spin liquid \cite{He2022}. Furthermore, impurities in topological systems can lead to the formation of topological polarons, which probe the underlying topology of the many-body state, e.g.~in fractional quantum Hall states  \cite{Grusdt2016}, Chern insulators \cite{Camacho2019} and one-dimensional symmetry protected topological phases \cite{Grusdt2019}. In a quantum Hall nematic, impurities lead to the formation of an excitonic bound state which can be in principle be probed using scanning tunneling microscopy \cite{Tam2020}. Finally, quantum dots coupled to fractional quantum Hall edges can be used to probe the edge states \cite{Keeling2008,Wagner2019a,Wagner2019b}.

In this work, we study strong impurities in quantum Hall systems. The problem of weak impurities in quantum Hall systems has already been addressed theoretically since the early days of quantum Hall physics. In particular, exact diagonalization studies showed that charged impurities lead to charge oscillations in the quantum Hall fluid \cite{Rezayi1985,Zhang1985}. In these studies, the impurities are weak enough such that the gap of the quantum fluid does not close. Later work has further refined this picture \cite{Aristone1993,výborný2007responseincompressiblefractionalquantum,Grass2020}.

Viewing the impurity as forming a small disk of depletion in the Hall droplet, the problem of impurity bound states is also related to that of edge reconstruction, which has been studied in detail~\cite{MacDonald1990,MacDonald1993,Johnson1991,Kane1994,Kane1995}. At the edge of a confining potential, the quantum Hall fluid can split into spatially separated regions with different filling fractions. This is an electrostatic effect \cite{Chklovskii1992}. It was first described using Hartree-Fock calculations \cite{Chamon1994,Wan2002,Dempsey1993,Karlhede1996}, showing that the edge of an integer quantum Hall system can split into alternating stripes with filling factors $\nu=0$ and $\nu=1$. However, Hartree-Fock is limited to the study of integer quantum Hall edges. Exact diagonalization studies in the disk geometry \cite{Wan2002,Wan2003,Hu2008,Hu2009}, Monte-Carlo calculations \cite{Meir1994} as well as a study using composite fermions \cite{Joglekar2003} later confirmed that edge reconstruction also occurs for fractional quantum Hall edges. The edge reconstruction also occurs for spinful systems \cite{Zhang2013,Khanna2017}. Exact diagonalization only allows access to small system sizes, making it hard to extrapolate to the thermodynamic limit. Therefore, later studies performed a variational calculation \cite{Khanna2021,Khanna2022} where they studied the competition between different types of edge reconstruction of the fractional quantum Hall edge. Such edge reconstruction of fractional quantum Hall edges was also confirmed in experiment \cite{Sabo2017}. Finally, time-reversal symmetric topological insulators can also undergo edge reconstruction \cite{Wang2017}.

The impurities we study in the present work sit in between these two extreme limits --- the weak impurity limit which slightly deforms the quantum Hall fluid and the strong potential limit, where the entire quantum Hall state reconstructs. In this intermediate regime, the effect of the impurity is characterized by bindung an integer number of quasiholes. These quasiholes are anyons and have fractionalized charge and statistics. Depending on the strength of the impurity as well as the chemical potential, different numbers of quasiholes can bind to the impurity. This setup therefore allows us to study interactions between the anyons. Interacting  anyons have been the subject of previous theoretical work, showing that anyons can form superconductors \cite{Lee1989}, excitons \cite{Rashba1993} and molecules \cite{Munos2020,xu2025dynamicsclustersanyonsfractional,gattu2025molecularanyonsfractionalquantum}. Furthermore, it has been shown that anyons can be trapped by an impurity potential~\cite{Hu2008b}. In that case the problem was studied in the disk geometry, which has a boundary and therefore the filling factor is not exactly fixed due to the edge excitations that can accommodate charge. In the present work, we study the effect of trapping \textit{several} anyons with an impurity potential in a spherical geometry.

Our motivation for revisiting this problem stems from recent experiments demonstrating the presence of fractional Chern insulators in the flat bands of twisted MoTe$_2$ \cite{cai2023signatures,zeng2023integer,Park2023,Xu2023,park2024ferromagnetism,xu2024interplay}. In particular, local magnetrometry measurements indicate the presence of charged impurities in the system \cite{Redekop2024}. In GaAs heterostructures, the impurities are spatially separated from the 2D electrons, which can be accounted for by taking a weak impurity with charge $Ze$ where $Z\sim0.1$ \cite{Zhang1985}. However, the impurities in twisted MoTe$_2$ may be significantly stronger and this warrants a detailed study of the strong impurity regime $Z\sim1$. In addition, the impurities in twisted MoTe$_2$ can be studied using exciton spectroscopy, which is not possible in GaAs since the latter is an optically active material. 

In the following, we use exact diagonalization in the spherical geometry to study the binding of anyons to an impurity. We focus on the $\nu=1/3$ Laughlin state as the paradigmatic example of an Abelian quantum Hall state. Computing the overlaps of low-lying states in the exact diagonalization spectrum with states generated from root partitions satisfying a generalized exclusion rule allows for the physical interpretation of those states in terms of quasiholes and quasiparticles. Furthermore, we compute the spectral function which can be experimentally measured using scanning tunneling microscopy. Finally, we compute the shift in the exciton binding energy due to the binding of anyons.

\section{Methods}

In order to study the quantum Hall system, we perform exact diagonalization in the spherical geometry. The spherical geometry has the advantage of lacking a boundary so that no edge states spoil the low-energy spectrum of the impurity. We study $N_e$ spinless electrons in the lowest Landau level, in the presence of $N_\phi$ flux quanta. The lowest Landau level on the sphere has $N_\phi+1$ orbitals indexed by $L_z=-\frac{N_\phi}{2},\dots,\frac{N_\phi}{2}$. Due to the shift on the sphere \cite{Wen1992}, the Laughlin $\nu=1/3$ state occurs when $N_\phi=3(N_e-1)$. 

The electrons interact via the Coulomb interaction projected to the lowest Landau level
\begin{equation}
    H_\textrm{int}=\frac{1}{2}\sum_{m_i}\langle m_1,m_2|V|m_3,m_4\rangle c_{m_1}^\dagger c_{m_2}^\dagger c_{m_4}c_{m_3},
\end{equation}
where $c_m$ annihilates an electron the Landau orbital with $L_z=m$. In addition, the electrons experience an impurity potential due to an impurity of charge $Ze$ located at the north pole of the sphere. The repulsive impurity potential in this case is
\begin{equation}
    V_\textrm{imp}(r)=\frac{Ze^2}{4\pi\epsilon r}
\end{equation}
leads to a single-particle potential $U_m$ for the orbital with angular momentum $m$ (see Supplement for further details):
\begin{equation}
    H_\textrm{imp}=\sum_mU_mc^\dagger_mc_m.
\end{equation}
The total Hamiltonian $H=H_\textrm{int}+H_\textrm{imp}$ is diagonalized using the iterative Lanczos algorithm. In the following, all the eigenenergies will be provided in units of $e^2/(4\pi\epsilon \ell_B)$, where $\ell_B$ is the magnetic length. In the presence of the impurity, $L_z$ remains a good quantum number and can hence be used to label the eigenenergies, while $L^2$ is no longer a good quantum number. 

We will find it useful to label states by their root configuration, where ``1" labels an orbital occupied by an electron and ``0" labels an empty orbital \cite{Bernevig2008,Thomale2011}. The root configuration for the $\nu=1/3$ Laughlin state for $N_e=4$ is $|1001001001\rangle$, where the $L_z$ eigenvalues of the orbitals are increasing from left to right (the leftmost orbital corresponds to the north pole of the sphere). The root configuration for the Laughlin state satisfies the rule that there is exactly one electron in three consecutive orbitals. Root configurations that violate this rule indicate the presence of quasiholes (less than one electron in three consecutive orbitals) or quasiparticles (more than one electron in three consecutive orbitals). Quasiholes and quasiparticles are the charged excitations of a quantum Hall fluid and lead to a characteristic charge modulation \cite{Haldane1985}. From a given root configuration, the corresponding many-body state can be derived using the Jack polynomials \cite{Bernevig2008,Thomale2011}. 

\begin{figure}
    \centering
    \includegraphics[width=\linewidth]{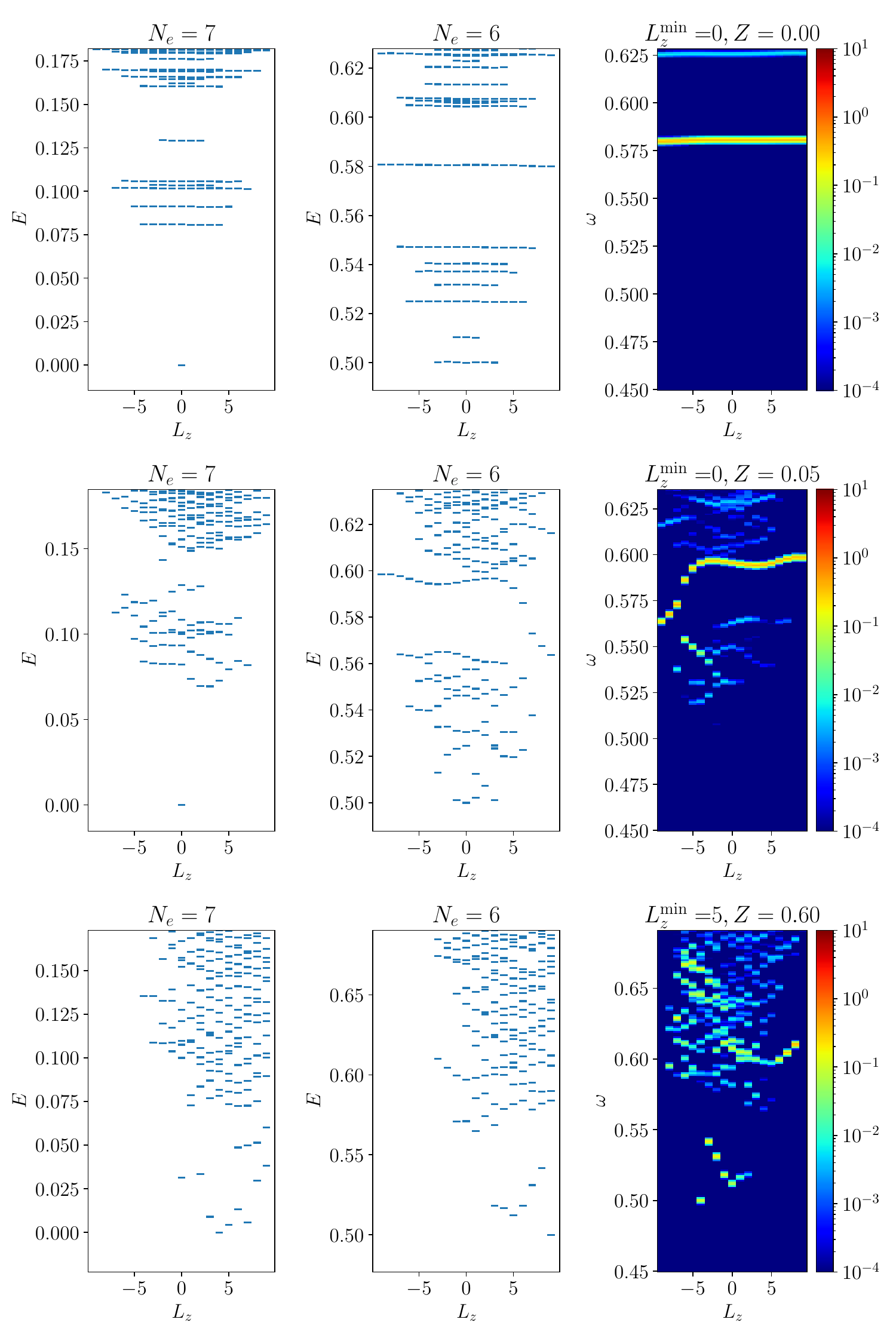}
    \caption{\textbf{Spectra and spectral function for $N_\phi=18$.} Top row: In the absence of the impurity ($Z=0$), the ground state is the $\nu=1/3$ Laughlin state and the excited states consist of the magnetoroton excitations. There is a large degeneracy coming from the $L^2$ quantum number. Middle row: For a weak impurity potential ($Z=0.05$), the spectral function is modified due to the presence of the impurity. Electrons are repelled from the impurity at the north pole, hence states with larger $L_z$ tend to have lower energy. The counting of the excited states gives information about the excitations of the Laughlin state \cite{Papic2018}. Bottom row: For a strong impurity potential  ($Z=0.6$), there is a transition in the ground state, which now occurs at finite $L_z$. The low-lying states in the spectrum at $N_e=6$ (which are reflected in the low-lying peaks in the spectral function) can be interpreted as different numbers of quasiholes bound to the impurity (see Fig.~\ref{fig:spectrum_Ne_6_Nphi_18}). The spectral function is computed with broadening of the delta function with $\delta E=0.01$.}
    \label{fig:Greens_fct_7_N}
\end{figure}

One way to probe the quantum Hall fluid is by scanning tunneling microscopy (STM). We ignore any disturbance on the state from the STM tip.  The STM can be used to measure the energy-resolved local density of states (spectral function), calculated as  
\begin{equation}
    \rho(\omega,L_z=m)=\sum_n|\langle N_e-1,n|c_m|N_e,0\rangle|^2\delta(\Delta E_n-\omega)
\end{equation}
where $|N_e,0\rangle$ is the ground state in the $N_e$ particle sector with energy $E_{N_e,0}$ and $|N_e-1,n\rangle$ is the $n$th excited state in the $N_e-1$ particle sector with energy $E_{N_e-1,n}$. We define the energy change from removing one electron as $\Delta E_n=E_{N_e-1,n}-E_{N_e,0}$. The spectral function has been proposed as a way to measure the composite fermion $\Lambda$ levels in fractional quantum Hall states \cite{Gattu2024,Pu2024} and it has also been proposed as a probe of fractional Chern insulating states such as twisted MoTe$_2$ \cite{Pichler2025}. Due to the relationship between angular momentum and position for the lowest Landau level orbitals, $\rho(\omega,m)$ will correspond to the local density of states in an annulus centered at radius $\sim \sqrt{2m}\ell_B$ around the impurity, where $\ell_B$ is the magnetic length. Measuring the spectral function around an impurity has been proposed as a way to identify and distinguish the Laughlin and Moore-Read states \cite{Papic2018}.

Besides the aforementioned STM, exciton spectroscopy can also be used for sensing bound anyons. \textcolor{black}{It is possible for the impurity potential to arise from the interactions between the electrons and an exciton. In that case, the internal structure is important as well. Such problems have been studied in the sphere geometry, where it is possible for an anyon to bind to the exciton, forming an ``anyon exciton" \cite{Wojs2000}. } We consider the following set-up (Fig.~\ref{fig:exciton_main_text} inset): We have a sensing bilayer which hosts interlayer excitons which are not mobile (e.g.~due to an impurity potential which pins them). The impurity potential is 
\begin{equation}
    V_\textrm{ex}(r)=\frac{e^2}{4\pi\epsilon}\bigg(\frac{1}{\sqrt{r^2+w^2}}-\frac{1}{\sqrt{r^2+(d+w)^2}}\bigg),
\end{equation}
where $d$ is the interlayer spacing of the sensing bilayer and $w$ is the separation between the sensing bilayer and the FQH sample. The change of the exciton binding energy from the effect of the FQH will be
\begin{equation}
\label{eq:delta_E_b}
    \Delta E_b=\int \mathrm{d}^2\vec{r}\ V_\textrm{ex}(r)\rho(r).
\end{equation}

\section{Results}

In Fig.~\ref{fig:Greens_fct_7_N} we show the spectra for $N_e=6,7$ particles as well as the spectral function. For $N_\phi=18$, the Laughlin $\nu=1/3$ state occurs at $N_e=7$. The top panel shows the results in the absence of an impurity potential. The ground state in the $N_e=7$ particle number sector is the Laughlin state. There are low-lying excitations which are well-separated from the rest of the spectrum; this is the neutral magnetoroton excitation. In the spectral function, the lowest lying states in the $N_e=6$ sector do not appear. The reason is that $L^2$ is a good quantum number absent the impurity potential and the states $c_m|\textrm{Laughlin}\rangle$ for different $m$ form an $L=\frac{N_\phi}{2}$ multiplet. The states $c_m|\textrm{Laughlin}\rangle$ thus have vanishing overlap with the low-lying states in the $N_e=6$ spectrum, which are multiplets with smaller $L^2$ quantum numbers. 

Let us first discuss the purely \emph{short-range} repulsive potential discussed in Ref.~\onlinecite{Rezayi1985}. In the limit where the impurity is infinitely strong, the north pole is completely unoccupied. The resulting ground state is well-described by projecting the Laughlin state, such that all the amplitudes with non-zero occupation of the north pole are projected out \cite{Rezayi1985}. The quantum Hall gap never closes in this case. A \emph{long-range} impurity  potential such as the one we study, in contrast, results in a gap closing and eventually in a transition in the ground state from an $L_z=0$ eigenstate to a state with $L_z>0$ as we will now  show.

The middle panel of Fig.~\ref{fig:Greens_fct_7_N} shows the spectrum in the presence of a weak long-range impurity potential, such that there was no gap closing when increasing $Z$ from $Z=0$. For a nonzero impurity potential, $L^2$ ceases being a good quantum number. The repulsive impurity is located at the north pole, i.e., centered at the orbital with $L_z=-9$. Therefore, states with higher total values of $L_z$, which have the electrons further from the impurity on average, are lower in energy. In the spectral function, this shows up in the slope of the main branch of the spectrum: It costs less energy to remove an electron from an orbital with small $L_z$, since such an electron is closer to the repulsive impurity. The other effect of the impurity potential is to render  low-lying states below the main branch of the spectrum in the spectral function visible, which are  hidden without impurity due to $L^2$ being a good quantum number. The counting of the states below and including the main branch ($1,1,2,3,4,\dots$ from left to right) serves as a fingerprint of the Laughlin state \cite{Papic2018}. 

\begin{figure}
    \centering
    \includegraphics[width=\linewidth]{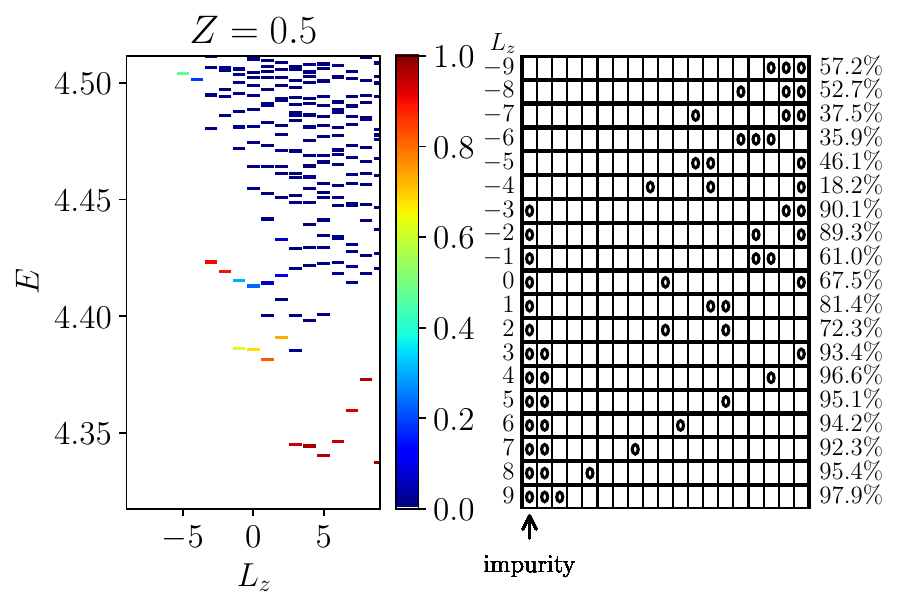}
    \caption{\textbf{Three quasihole spectrum ($N_e=6$, $N_\phi=18$).} The colorbar indicates the overlap with the root configuration states labelled on the right. The lowest energy state has two quasiholes bound to the impurity and one quasihole released. The branch of excited states consists of the dispersion of the additional quasihole that has been released. In addition, the state at $L_z=9$ consists of three quasiholes bound to the impurity. Impurity strength $Z=0.5$. }
    \label{fig:spectrum_Ne_6_Nphi_18}
\end{figure}

\begin{subequations}
Finally, for even stronger long-range impurity potentials (bottom panel of Fig.~\ref{fig:Greens_fct_7_N}), there is a transition in the ground state. The ground state in the $N_e=7$ sector is now at finite $L_z=4$. In order to interpret the spectral function in this case, we need to understand the energy spectrum in the $N_e-1$ particle sector. The root partitions for the $N_e-1$ sector (here $N_e=5$ as an example) are 
\begin{align}
    L_z=0&: \underset{\circ}{0}10010010\underset{\circ}{0}\underset{\circ}{0}01,\\
    L_z=1&: \underset{\circ}{0}10010\underset{\circ}{0}010\underset{\circ}{0}01,\\
    L_z=2&: \underset{\circ}{0}\underset{\circ}{0}1001001001\underset{\circ}{0},\\
    L_z=3&: \underset{\circ}{0}\underset{\circ}{0}10010010\underset{\circ}{0}01,\\
    L_z=4&: \underset{\circ}{0}\underset{\circ}{0}10010\underset{\circ}{0}01001,\\
    L_z=5&: \underset{\circ}{0}\underset{\circ}{0}10\underset{\circ}{0}01001001,\\
    L_z=6&: \underset{\circ}{0}\underset{\circ}{0}\underset{\circ}{0}1001001001.
\end{align}
These root partitions all satisfy the rule that there is not more than one electron in three consecutive orbitals and thus they are low-energy states. The notation $\underset{\circ}{0}$ denotes the position of a quasihole. The root partitions thus describe states with one ($L_z=0,1$), two ($L_z=2,3,4,5$) or three ($L_z=6$) quasiholes bound to the impurity. 
\end{subequations}

In order to substantiate this interpretation of the spectrum, we compute overlaps of the exact diagonalization states with the states generated from certain root partitions satisfying the Laughlin exclusion rules (not more than one electron in three orbitals). In Fig.~\ref{fig:spectrum_Ne_6_Nphi_18} we show the overlaps in the $N_e=6, N_\phi=18$ sector. The root partitions listed are those root partitions satisfying the exclusion rule that have the highest overlap with the lowest energy state in a given $L_z$ sector. In this particle number sector, there are three additional quasiholes relative to the Laughlin state. There are two competing energy scales in the problem: On the one hand, the quasiholes are charged and therefore repel one another. On the other hand, the repulsive potential for the electrons results in a single-particle potential for the quasiholes attracting them to the impurity, which leads the quasiholes to preferentially occupy orbitals with smaller $L_z$. The impurity can bind $0,1,2,3$ quasiholes in the parameter range we studied. There is a jump in the lowest energy state as a function of $L_z$ every time an additional quasihole gets bound to the impurity.  The lowest energy state has two quasiholes bound to the impurity, however the state with three quasiholes bound is closely competing.

\begin{figure}
    \centering
    \includegraphics[width=\linewidth]{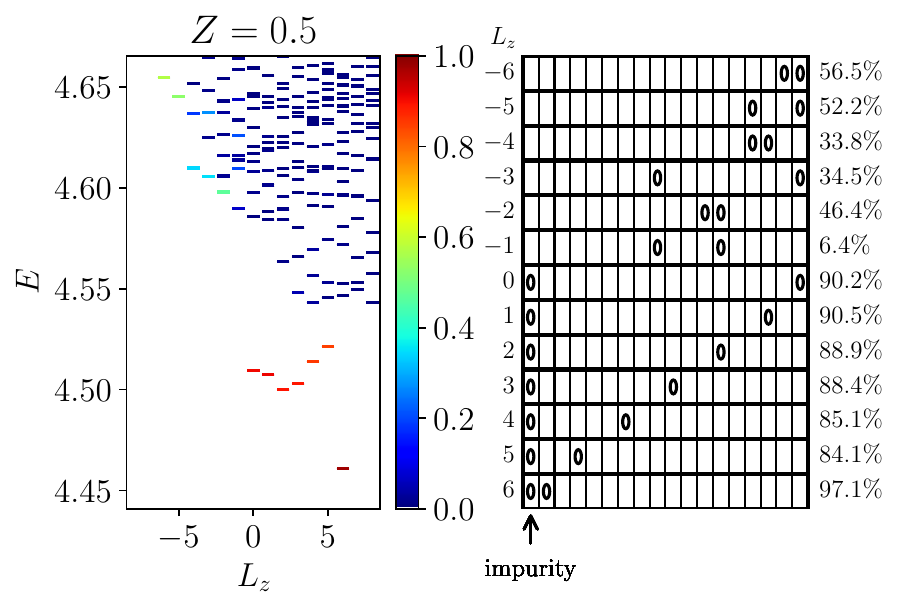}
    \caption{\textbf{Two quasihole spectrum ($N_e=6$, $N_\phi=17$).} The colorbar indicates the overlap with the root configuration states labelled on the right. The lowest energy state has two quasiholes bound to the impurity. The branch of excited states consists of a single quasihole bound to the impurity and an additional quasihole that has been released. Impurity strength $Z=0.5$.}
    \label{fig:spectrum_Ne_6_Nphi_17}
\end{figure}

Next, in Fig.~\ref{fig:spectrum_Ne_6_Nphi_17} we present results for the case with one orbital less, i.e.~$N_\phi=17$, which corresponds to removing a quasihohle from the system. In this case the impurity can bind $0,1,2$ quasiholes. The lowest energy state has two quasiholes bound. There are excited states, where one quasihole has been released and only one quasihole remains bound to the impurity. The energy difference between the lowest energy state and the brach of low-lying excitations is the ionization energy of the two quasihole state. 

\begin{figure}[t]
    \centering
    \includegraphics[width=\linewidth]{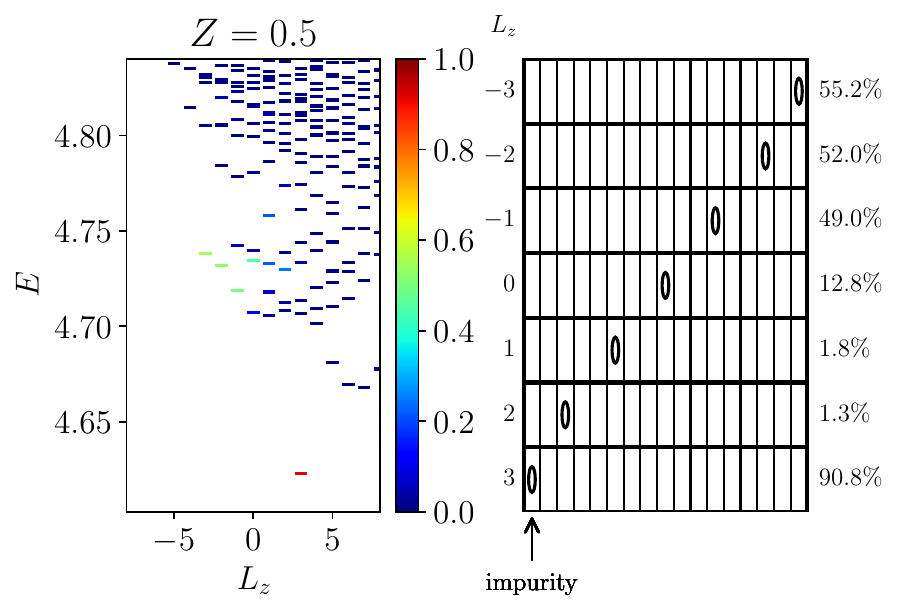}
    \caption{\textbf{One quasihole spectrum ($N_e=6$, $N_\phi=16$).} The colorbar indicates the overlap with the root configuration states labelled on the right. The lowest energy state has one quasihole bound to the impurity. The branch of excited states consists of two quasiholes bound to the impurity and an additional quasiparticle that has been released.  Impurity strength $Z=0.5$.}
    \label{fig:spectrum_Ne_6_Nphi_16}
\end{figure}

Finally, in Fig.~\ref{fig:spectrum_Ne_6_Nphi_16} we consider the case with $N_\phi=16$, such that there is only one additional quasihole relative to the Laughlin state filling. This single quasihole remains bound to the impurity in the ground state of the system. However there are excited states, where two quasiholes are bound to the impurity and there is an additional quasiparticle released. 

External probes, such as STM, will always probe transitions between eigenstates of the system that differ by entire electrons, which corresponds to  three quasiholes to be added or removed at a time. However, the ground state in presence of the impurity may have any number of quasiholes bound to the impurity. For meaningful comparison of energies between sectors with different numbers of quasiholes, we must compare sectors with different numbers of flux quanta for a fixed number of electrons $N_e$. 

\begin{figure}
    \centering
    \includegraphics[width=\linewidth]{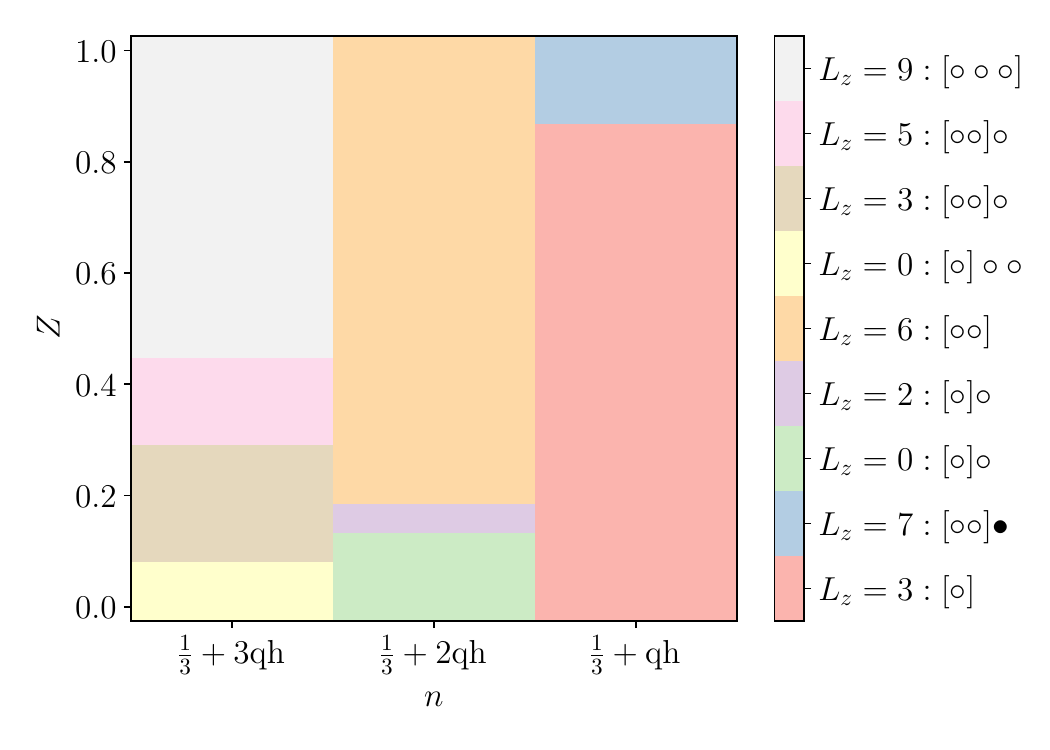}
    \caption{\textbf{Phase diagram for different quasihole sectors ($N_e=6)$.} We compute the ground state energy in different $N_\phi$ sectors in the presence of an impurity potential of charge $Z$. We show the representative root configuration for each state. $\circ$ ($\bullet$) denotes the position of a quasihole (quasiparticle). \textcolor{black}{The square brackets indicate quasiholes bound to the impurity. The difference between the $L_z=3$ and $L_z=5$ states for $\frac{1}{3}+3$qh is in the separation of the unbound quasihole from the impurity. For $L_z=5$, the unbound quasihole is closer to the impurity.}}
    \label{fig:phase_diagram}
\end{figure}

The corresponding ground state as a function of impurity strength and density is shown in Fig.~\ref{fig:phase_diagram}. As the strength of the repulsive impurity $Z$ is increased, the number of quasiholes bound to the impurity increases, as the impurity potential overcomes the mutual repulsion of the quasiholes. The impurity can also generate quasiparticle-quasihole pairs, with the quasiholes bound to the impurity. 


\textcolor{black}{In order to probe the different quasihole sectors, the density can be tuned via gating. It is possible to either apply a gate potential locally via an STM tip, or by gating the entire sample, while ensuring one remains within the quantum Hall plateau.}

\begin{figure}
    \centering
    \includegraphics[width=\linewidth]{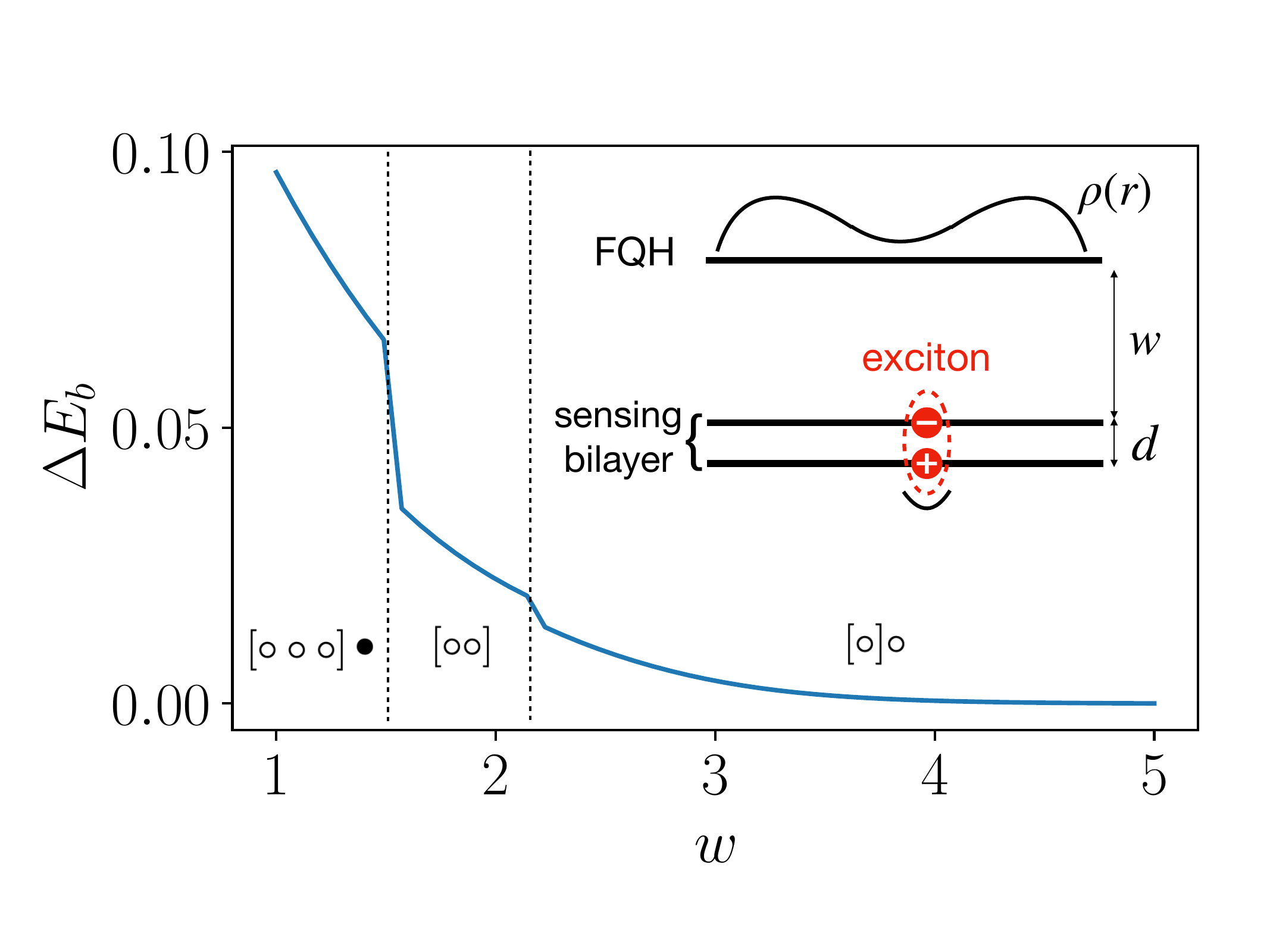}
    \caption{\textbf{Energy shift of the exciton binding energy due to anyon capture.} Inset: We consider an experimental setup with a pinned interlayer exciton in a sensing bilayer with interlayer separation $d$. The sensing bilayer is separated by distance $w$ from the FQH state. The exciton acts as an impurity on the FQH state, inducing a modulation of the FQH density $\rho(\vec{r})$, which influences the exciton binding energy. Main panel: The exciton binding energy in the sensing layer $\Delta E_b$ changes depending on the distance $w$ between the exciton and the FQH electrons. The exciton binding energy jumps each time an anyon is captured. $N_e=6,\ N_\phi=17$, $d=0.8\ell_B$}
    \label{fig:exciton_main_text}
\end{figure}

In Fig.~\ref{fig:exciton_main_text} (main panel), we show  the change in the exciton binding energy $\Delta E_b$ due to the interaction with the anyons. Between the states with different number anyons bound, the binding energy of the exciton changes discontinuously. We use $w$ as a tuning parameter allowing us to access states with different numbers of anyons bound to the exciton. Experimentally, the density is another possible tuning parameter, which can also be used to tune the number of bound anyons as shown in Fig.~\ref{fig:phase_diagram}.  Thus, performing an optical experiment as a function of doping should reveal several shifted exciton peaks.

In practice, one may also consider the setup in the absence of a magnetic field, where we have an FCI state instead of an FQH state. For example, the FCI state could exist in twisted MoTe$_2$ and the sensing layer would consist of a heterobilayer of transition metal dichalcogenides with interlayer distance $d$, separated from the FCI by a slab of hexagonal boron nitride of thickness $w$.

\section{Conclusion}

We have shown that impurities in a quantum Hall fluid can bind a varying number of quasiholes. The number of quasiholes bound to the impurity will depend on the impurity strength as well as on the chemical potential of the system. Scanning tunneling microscopy can be used to image the charge distribution around an impurity in a quantum Hall fluid, and such experiments have been performed successfully \cite{hu2024highresolutiontunnelingspectroscopyfractional,Luican2014,chiu2024highspatialresolutioncharge}. We have shown that there is a transition in the ground state as a function of the chemical potential. For a sufficiently strong impurity, within a single plateau one can scan through several states of the impurity via the back gate or by varying the magnetic field. 

 The strength of the impurity potential will be intimately related with the nature of the impurity. In GaAs heterostructures, the impurity may reside in a layer adjacent to the quantum Hall fluid itself, the impurity strength is correspondingly reduced. For the case of a strong impurity which is the main focus of our work, this traditional platform may therefore be unsuitable to access the regime of interest. The twisted transition metal dichalcogenide MoTe$_2$ has shown FCI at many filling fractions including Laughlin states and states of the Jain sequence \cite{cai2023signatures,zeng2023integer,Park2023,Xu2023,park2024ferromagnetism,xu2024interplay}. MoTe$_2$ is known to have impurities \cite{Redekop2024}, and since these may reside in the same layer as the FCI, we would expect them to be in the strong impurity regime studied here. It would therefore be of great interest to perform STM imaging or exciton spectroscopy on impurities in twisted MoTe$_2$ in the FCI regime. Theoretically, impurities in twisted MoTe$_2$ have been studied in Ref.~\cite{liu2025characterizationfractionalcherninsulator}.

In future work, it would be interesting to study the strong-impurity regime for the case of non-Abelian states such as the Moore-Read state, or for fractional topological insulators which may be relevant in twisted MoTe$_2$ \cite{kwan2024abelianfractionaltopologicalinsulators}. Furthermore, it would be of interest to generalize the calculation to the fractional Chern insulator regime, where lattice effects are relevant.

\textit{Note added:} During completion of this work we became aware of related work by Mostaan, Goldman, İmamoğlu, Grusdt, which has been submitted in the same arxiv posting.

\section{Acknowledgments}

We thank Ataç İmamoğlu, Tomasz Smole\'nski and Ajit Srivastava for useful discussions. Exact diagonalization calculations were performed using DiagHam. G.W.~is supported by the Swiss National Science Foundation (SNSF) via Ambizione grant number PZ00P2-216183. T.N. acknowledges support from the Swiss National Science Foundation through a Consolidator Grant (iTQC, TMCG-2\_213805). 

\bibliographystyle{unsrtnat}
\bibliography{bib}

\newpage
\clearpage
\begin{appendix}
\onecolumngrid
	\begin{center}
		\textbf{\large --- Supplementary Material ---\\Sensing the binding and unbinding of anyons at impurities}\\
		\medskip
		\text{Glenn Wagner, Titus Neupert}
	\end{center}
	
		\setcounter{equation}{0}
	\setcounter{figure}{0}
	\setcounter{table}{0}
	\setcounter{page}{1}
	\makeatletter
	\renewcommand{\theequation}{S\arabic{equation}}
	\renewcommand{\thefigure}{S\arabic{figure}}
	\renewcommand{\bibnumfmt}[1]{[S#1]}

\section{Exact diagonalization in the sphere geometry}

We consider the electrons on the surface of a sphere threaded by flux $N_\phi=2q$, often known as the Haldane sphere \cite{Haldane1983}.  The radius of the sphere is $R=\sqrt{q}\ell_B$, where $\ell_B$ is the magnetic length. The single-particle wavefunctions of the electrons are given by the monopole harmonics \cite{WU1976,Wu1977}. In the lowest Landau level, the electron orbitals are labelled by their $L_z$ eigenvalue $m=-q,\dots,+q$. They take the form
\begin{equation}
    Y_{qm}(\theta,\phi)=\sqrt{\frac{2q+1}{4\pi}\binom{2q}{q-m}}(-1)^{q-m}v^{q-m}u^{q+m},
\end{equation}
where $u,v$ are the so-called spinor coordinates
\begin{equation}
    u=\cos(\theta/2)e^{i\phi/2}, v=\sin(\theta/2)e^{-i\phi/2}
\end{equation}
and $(\theta,\phi)$ are the usual spherical coordinates on the sphere. We now consider an impurity of charge $Ze$ at the north pole of the sphere. This induces a single-particle potential
\begin{equation}
    U(\theta)=\frac{Ze^2}{4\pi\epsilon}\frac{1}{2R|\sin\theta/2|},
\end{equation}
where $2R|\sin\theta/2|$ is the chord distance to the north pole. The leads to a one-body-term for the orbitals of the form
\begin{equation}
\label{eq:U_m}
    U_m=\int_0^{2\pi}\mathrm{d}\phi \int_0^{\pi}\mathrm{d}\theta\sin\theta \ U(\theta) |Y_{qm}(\theta,\phi)|^2
\end{equation}
leading to a single-particle term in the Hamiltonian
\begin{equation}
    H_\textrm{imp}=\sum_mU_mc^\dagger_mc_m.
\end{equation}
The interaction term takes the form
\begin{equation}
    H_\textrm{int}=\frac{1}{2}\sum_{m_i}\langle m_1,m_2|V|m_3,m_4\rangle c_{m_1}^\dagger c_{m_2}^\dagger c_{m_4}c_{m_3},
\end{equation}
where the matrix elements are 
\begin{equation}
    \langle m_1,m_2|V|m_3,m_4\rangle=\sum_{L=0}^{2q}\sum_{M=-L}^L\langle qm_1,qm_2|L,M\rangle V_L\langle L,M|qm_3,qm_4\rangle.
\end{equation}
$\langle L,M|qm_1,qm_2\rangle$ are Clebsch-Gordan coefficients and $V_L$ are known as the Haldane pseudopotentials, which specify the energy penalty for electrons with relative angular momentum $L$. See Ref.~\cite{fano1986configuration} for an explicit expression for $V_L$ for the case of the Coulomb interaction.

We now consider the following set-up (Fig.~\ref{fig:exciton_main_text}a): We have a sensing bilayer which hosts interlayer excitons which are not mobile (e.g.~due to an impurity potential which pins them). The impurity potential is 
\begin{equation}
    V(r)=\frac{e^2}{4\pi\epsilon}\bigg(\frac{1}{\sqrt{r^2+w^2}}-\frac{1}{\sqrt{r^2+(d+w)^2}}\bigg).
\end{equation}
Translating to the sphere geometry, we have 
\begin{equation}
    U(\theta)=\frac{e^2}{4\pi\epsilon}\bigg(\frac{1}{\sqrt{(2R\sin\theta/2)^2+w^2}}-\frac{1}{\sqrt{(2R\sin\theta/2)^2+(d+w)^2}}\bigg)
\end{equation}
and again, the single-particle term can be computed using Eq.~\eqref{eq:U_m}. The change to the exciton binding energy on the sphere (Eq.~\ref{eq:delta_E_b}) is
\begin{equation}
    \Delta E_b=2\pi\int \mathrm{d}\theta\ \sin\theta \ U(\theta)\rho(\theta).
\end{equation}

\end{appendix}

\end{document}